\documentclass[10pt]{article}

\usepackage{amsmath}
\usepackage{amssymb}

\usepackage{graphicx}
 \graphicspath{{./figures/}}

\usepackage{cite}
\usepackage{color} 
\usepackage{bm} 
\usepackage{lineno}

\usepackage[labelfont=bf,labelsep=period,justification=raggedright]{caption}

\makeatletter
\renewcommand{\@biblabel}[1]{\quad#1.}
\makeatother

\date{}

\newcommand{\citet}[1]{\cite{#1}}
\newcommand{\citep}[1]{\cite{#1}}
\newcommand{\tr}{^{\sf T}}
 \newcommand{\ve}[1]{\bm{#1}}
 \newcommand{\ven}[1]{\bm{\mathit{#1}}}
 \newcommand{\ma}[1]{\mathbf{#1}}
 \newcommand{\dd}{{\rm d}}
 \newcommand{\nn}{\nonumber}
 \newcommand{\ee}{\text{e}}
 \newcommand{\panel}[1]{\textsf{\textbf{#1}}}
\usepackage[normalem]{ulem}

\begin{document}

\begin{flushleft}
{\Large
\textbf{How Levins' dynamics emerges from a Ricker metapopulation model on the brink of extinction}
}
\\
F. {El\'ias-Wolff}$^{1}$, 
A. Eriksson$^{2}$, 
A. Manica$^{2}$
B. Mehlig$^{1,3,\ast}$, 
\\
\bf{1} Department of Physics, University of Gothenburg, SE-41296 Gothenburg, Sweden
\\
\bf{2}  Department of Zoology, University of Cambridge, Cambridge, CB2 3EJ, UK
\\
\bf{3} The Linnaeus Centre for Marine Evolutionary Biology, University of Gothenburg, SE-405 30 Gothenburg, Sweden
\\
$\ast$ E-mail: Bernhard.Mehlig@physics.gu.se      
\end{flushleft}


\section*{Abstract}
Understanding the dynamics of metapopulations close to extinction is of vital importance for management.  Levins-like models, in which local patches are treated as either occupied or empty, have been used extensively for this purpose, but they ignore the important role of local population dynamics. In this paper, we consider a stochastic metapopulation model where local populations follow a Ricker dynamics, and use this framework to investigate the behaviour of the metapopulation at the brink of extinction. As long as dispersal rates are not too large, the system is shown to have a time evolution consistent with Levins' dynamics.  We derive analytical expressions for the colonisation and extinction rates ($c$ and $e$) in Levins-type models in terms of reproduction, survival, and dispersal parameters of the local populations, providing an avenue to parameterising Levins-like models from the type of information on local demography that is available for a number of species. To facilitate applying our results, we provide a numerical implementation for computing $c$ and $e$.\\
{\em Keywords}: colonisation, extinction, Levins' equation, Ricker model

\section*{Introduction}
Most species live in systems of populations connected by dispersal, also known as metapopulations. While small local populations can go extinct frequently (due to stochastic events, competition, predation and other factors), the whole metapopulation persists due to dispersal allowing for the recolonisation of empty patches and the rescuing of patches close to local extinction.

Levins \citet{Levins1969} provided the first mathematical representation of metapopulations. His approach takes the simplifying assumption of treating individual patches as either occupied or empty, thus ignoring the local patch dynamics. The total fraction $p$ of occupied patches changes as a function of time, and is described by Levins' equation:
\begin{equation}\label{eq:levins}
    \frac{\dd p}{\dd t} = c p(1-p) - e p \, .
\end{equation}
Here $c$ is the rate (per patch) of successful colonisation of empty patches, and $e$ is the corresponding extinction rate.
Levins' equation has formed the basis of most metapopulation modeling, and the basic model has been extended to account 
for a large number of additional factors, such as allowing for different patch sizes (so-called core-satellite models), 
Allee effects, and multi-species interactions.  

Despite the extensive literature on the behaviour of Levins-like models 
(summarised e.g. by Etienne \citet{Etienne2002}),
a fundamental question has received limited attention:
under which conditions, that is to say for which type of local dynamics 
do real metapopulations exhibit a Levins-like behaviour? 
Keeling \citet{Keeling2002} investigated numerically a metapopulation dynamics based on local coupled maps, in order to determine whether Levins' model is a good approximation. 
He found that for certain parameter values, the results of numerical simulations of the metapopulation fitted well to the form of Levins' equation 
(similar results were obtained in \citet{Wennberg12}).
Fronhofer {\em et al.} \citet{Fronhofer2012} have performed numerical simulations of a spatially structured population model, to answer the question under which conditions 
it behaves qualitatively as a metapopulation, Eq.~(\ref{eq:levins}), but make no attempt at determining the rates $c$ and $e$.
Fronhofer {\em et al.} \citet{Fronhofer2012} 
base their conclusions on a number of criteria (e.g. levels of patch occupancy and turnover) and conclude that on the brink
of extinction the dynamics of their model is goverend by a Levins-like balance between patch colonisations and extinctions.

Even  for conditions under which metapopulations can be shown to follow a Levins-like behaviour \citep{Hanski1995}, it remains a challenge to parameterise the corresponding colonisation and extinction rates, as it requires long time series of censuses of a number of patches. 
While it is straightforward to generate such data from simulations, they are hard to obtain for empirical systems; indeed,
this type of comprehensive data is rare in the ecological literature. See Ref. \citep{Hanski1995} for a study of an endangered butterfly population shown to behave Levins-like.

A possible solution to this problem is to develop an analytical formulation of the local stochastic dynamics within a metapopulation framework, thus providing a direct link between local demographic parameters and the colonisation and extinction rates of patches at the metapopulation level. Using such a framework,
Eriksson {\em et al.} \citet{Eriksson2011} estimated metapopulation extinction times for the case of local dynamics with time-continuous birth-death dynamics.
In a related approach, Lande {\em et al.} \citet{Lande1998} 
have proposed an elegant stochastic generalisation of Le\-vins' approach for
continuous population dynamics, in order to quantify how the local patch dynamics 
affects the properties of the quasi-steady state of the metapopulation.

However many population models commonly used in the literature are based
on discrete time evolution in terms of non-overlapping generations of individuals. 
In such models, the dynamics is usually represented by a collection of coupled maps \citep{Ylikarjula2000,Ranta1995,Wysham2008}. 
Examples are the logistic map, the Ricker map \citep{Ricker1954}, and the Hassell map \citep{Hassell1975}.

In the following we develop an analytical formulation of a metapopulation where the local population dynamics is governed by a stochastic version of the Ricker map. 
We mention that the spatial structure of metapopulations with local Ricker dynamics was investigated 
in {Refs.~\citep{Hastings1994,Ruxton1996,Dey2006}.

In this paper we investigate under which conditions a metapopulation model based on Ricker's map exhibits a Levins-like behaviour. 
We then estimate patch colonisation and extinction rates (the two key parameters in Levins' equation) in terms of reproduction, survival and dispersal parameters of the local populations.

The remainder of this paper is organised as follows.
We first describe the metapopulation model. 
Then we derive a deterministic law for the evolution
of this population, in the limit of many patches. 
We then proceed to show how this dynamics simplifies when
the population is close to extinction, and how  a discrete
version of Levins' equation (\ref{eq:levins})
is obtained. Finally, in a concluding section, we discuss
our results and put them in context.
Mathematical details are deferred to appendices.

\section*{Stochastic metapopulation model}
\label{sec:model}
Our model consists of a population distributed among $N$ patches evolving in discrete, non-overlapping generations. 
It is closely related to the model studied numerically by Keeling \citet{Keeling2002}. 
During each generation, three processes take place, in this order: reproduction, survival to maturity, and dispersal.
The first two phases occur within patches, and are described by a stochastic Ricker map. 
The Ricker map as it is commonly used corresponds to the deterministic dynamics:
 \begin{equation}\label{eq:ricker}
 \eta_{t + 1} = \eta_{t} R \ee^{-\alpha \eta_{t}} \,.
 \end{equation}
Here $\eta_t$ denotes the number of adults in a local population in generation $t$, 
the parameter $R$ characterises the average offspring size, that is, the fecundity rate. The parameter $\alpha$ determines the probability of density-dependent survival (such as the incidence rate of cannibalism by adults). 

Stochasticity is added in the following manner. Consider a patch with $\eta_t$ individuals in generation $t$. 
We take the number of individuals after reproduction to be Poisson--distributed with mean $R \eta_t$. 
During the second phase, individuals survive to maturity independently with probability $\exp(-\alpha \eta_t)$. Thus the number of adults 
$\eta_{t+1}$ in the next generation is a random variable with mean equal to the right-hand side of Eq.~\eqref{eq:ricker}. 
During the third phase, the dispersal phase, individuals independently move away from the patch with probability $m$. 
Finally, the migrants from all patches are gathered in a common pool from where
the target patch of each migrant is independently and randomly chosen, and thus 
the migrants are uniformly randomly distributed among all $N$ patches. Fig.~\ref{fig:1} 
illustrates this migration process. 
The pseudocode for the direct numerical simulations of the model described here is given in algorithm 1.
Note that because we start from a Poisson distributed number of offspring and use binomial sampling in the survival and dispersal phases, 
it follows that the patch population sizes are Poisson distributed in each phase \cite{Karlin1998}
(and that the population size is statistically independent between patches, conditional on the patch population sizes in the previous generation
).
\begin{figure}[t]
   \centering
   \includegraphics[width=0.7\textwidth]{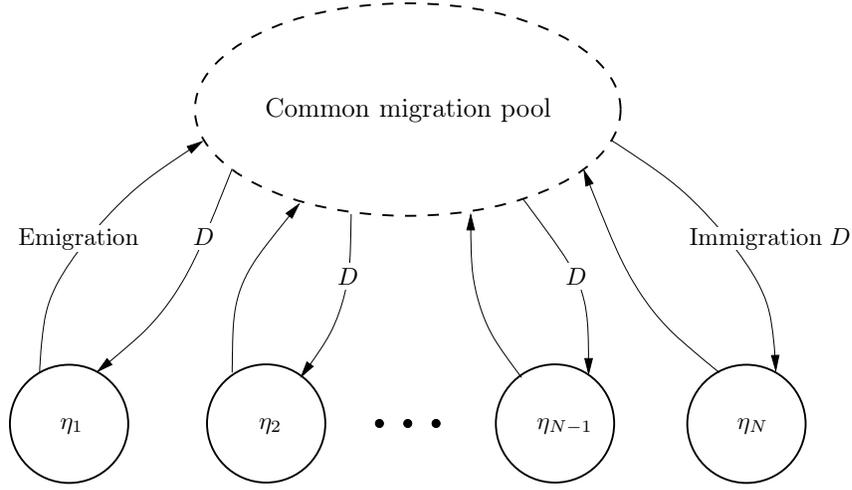}
   \caption{\label{fig:1} Illustration of the migration scheme in the metapopulation model. 
   The metapopulation consists of $N$ patches $j$ with population sizes $\eta_j$. 
   In each generation, after reproduction and survival to maturity, 
   a random number of emigrants leaves each patch for  the
   common pool. All of these individuals are then redistributed to the patches. The expected
   number of immigrants per patch has mean value $D(t)$, see Eq.~\eqref{eq:It} in appendix~A. 
}
\end{figure}

\section*{Dynamics in the limit of infinitely many patches}
\label{sec:detdyn} 
In this section we analyse the metapopulation dynamics in the limit of infinitely many patches.
Since all patches are equivalent (the local population dynamics
is given by the parameters $\alpha$ and $R$), it is convenient to describe
the state of the metapopulation by recording the frequencies of patches
with a given number of individuals. In other words, the state
of the metapopulation is described by the vector $(n_0, n_1, \dotsc)$, where $n_0$ denotes the number of empty patches, $n_1$ the number of patches containing one individual, etc.
In the limit of $N\rightarrow \infty$ we expect the frequencies $f_j = n_j / N$ to approach $N$-independent limits,
and change deterministically from generation $t$ to $t+1$:
\begin{equation}\label{eq:detdyn1}
f_{i}(t+1) = \sum_{j=0}^{\infty} P_{j \to i}(t) f_{j}(t) \, .
\end{equation}
Here $P_{j \to i}(t)$ is the probability that a patch inhabited by $j$ individuals in generation $t$ has population size $i$ in generation $t+1$.
Since the fractions $f_i$ must sum to unity, it is convenient characterise the state of the metapopulation using the vector $\ve f(t)$
with components $(f_1(t),f_2(t),\ldots)$.  Note that $P_{j \to i}(t)$ depends on $\ve f(t)$. This dependence is determined by the migration scheme.
In appendix~A we derive the explicit form of $P_{j \to i}$. For the following discussion the details do not matter.

The long-term behaviour of Eq.~\eqref{eq:detdyn1} is decided by the values of the parameters $R$, $\alpha$, and by the value of the emigration rate $m$. 
For sufficiently large emigration rates, the metapopulation converges to a quasi-steady state, described 
by the steady-state solution $\ve f^\ast$ of the dynamics \eqref{eq:detdyn1}.
In a finite metapopulation, the state cannot persist {\em ad infinitum} 
since finite populations must eventually become extinct unless they continue to grow \citep{Jagers}. 
It is nevertheless of interest to study the limit $N\rightarrow\infty$ because
stable steady states in this limit correspond to very long-lived, quasi-steady states in finite but large populations.
\begin{figure}[tp]
   \includegraphics[width=0.24\textwidth,height=0.24\textwidth]{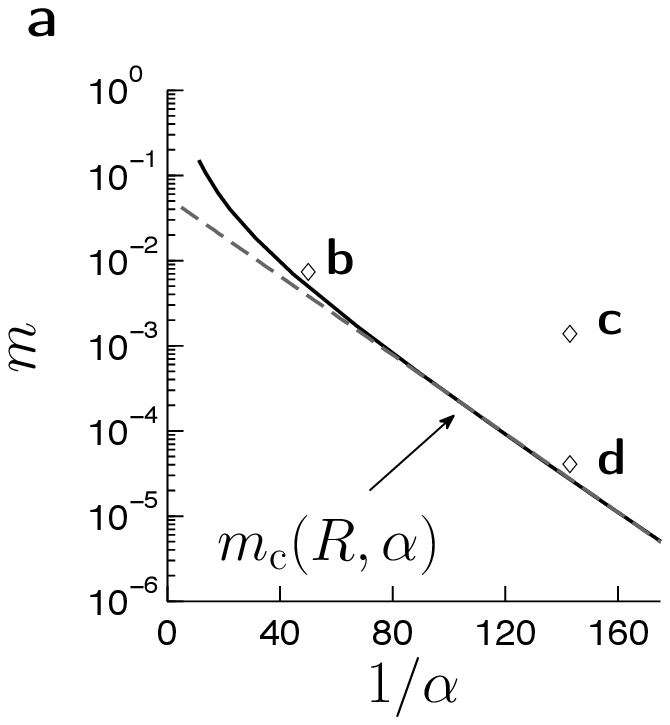}
   \includegraphics[width=0.24\textwidth,height=0.24\textwidth]{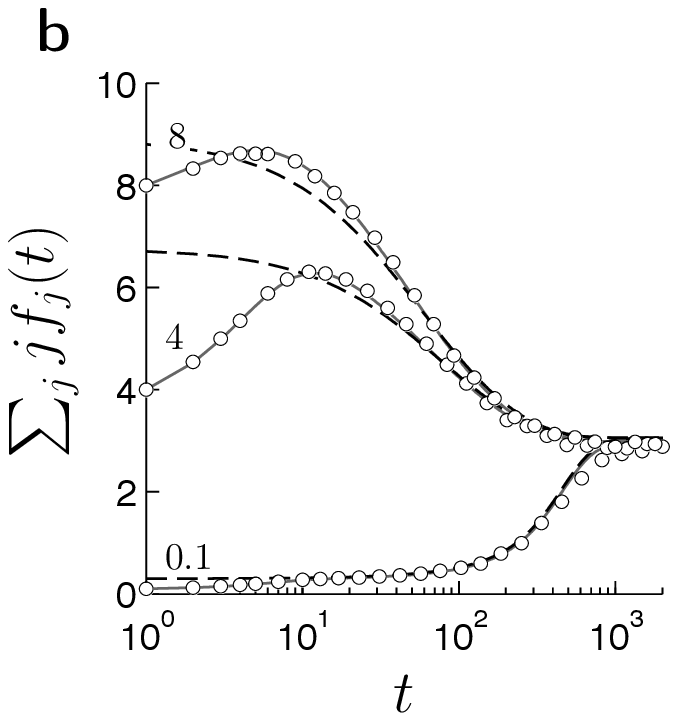}
   \includegraphics[width=0.24\textwidth,height=0.24\textwidth]{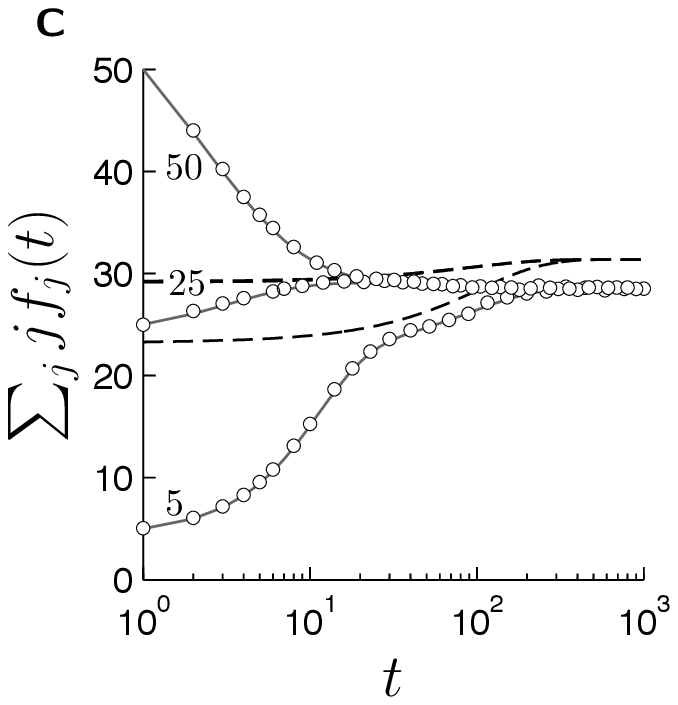}
   \includegraphics[width=0.24\textwidth,height=0.24\textwidth]{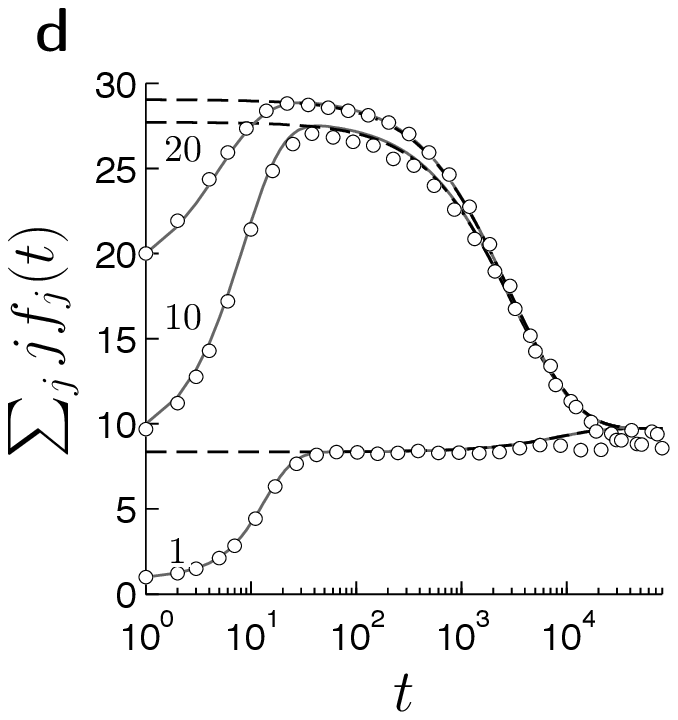}
   \caption{\label{fig:2}
(\textsf{\textbf{a}}) Line in the $\alpha^{-1}$-$m$ plane
distinguishing  metapopulation persistence (in the limit of $N\rightarrow\infty$) from extinction, for fecundity $R = 1.25$. 
Symbols ($\Diamond$) indicate the parameters values used in the remaining panels
of this figure. The dashed line corresponds to an exponential law $m_{\rm c} \sim \exp(-S / \alpha)$ with $S\approx 0.0532$.
(\textsf{\textbf{b}}-\textsf{\textbf{d}}) Comparison
between the deterministic dynamics (\ref{eq:detdyn1}), solid lines,  and the results of direct numerical
simulations of the metapopulation model (symbols) for  $N=250$ (panel \textsf{\textbf{b}}) and $N=100$ (panels \textsf{\textbf{c}} and \textsf{\textbf{d}}) patches.
Other parameter values as indicated in panel \textsf{\textbf{a}}.  Shown is the average patch population size $\sum_j jf_j$ as a function of time.
Initial condition: Poisson distributed $f_{j}$ with mean given in the figure.
The direct numerical simulations
were averaged over $100$ independent runs. The dashed lines correspond to Levins' dynamics (\ref{eq:discr_lev}-\ref{eq:levrates}). }
\end{figure}

As the emigration rate $m$ decreases (at fixed values of  $R$ and $\alpha$), the infinitely large metapopulation approaches the brink of extinction: 
there is a critical value $m_{\rm c}$ where the infinite metapopulation becomes unstable and ceases to persist.
For $m < m_{\rm c}$, the steady state $\ve f=\ven 0$ is stable, thus extinction is certain. 

How does $m_{\rm c}$ depend on the life-history parameters $R$ and $\alpha$?
In order to answer this question we must investigate the linear stability of $\ve f^\ast$ by linearising the dynamical equation (\ref{eq:detdyn1}).
This procedure is briefly described in appendix~B. The result of the analysis is shown in Fig.~\ref{fig:2}{\panel a} 
for $R=1.25$. We see that $m_{\rm c} \sim \exp(-S(R)/\alpha)$ for small values of $\alpha$.
The line separating stable from unstable steady states is shown as a solid line in the $\alpha^{-1}$-$m$-plane in Fig.~\ref{fig:2}{\panel a}.
Above this line the non-trivial steady states $\ve f^\ast$ are linearly stable (the infinite metapopulation persists {\em ad infinitum}), 
below this line they are unstable.  

Panels ${\panel b}$ to {\panel d} in Fig.~\ref{fig:2} show a comparison
between the dynamics obtained from Eq.~(\ref{eq:detdyn1}), solid lines,  and the results of direct stochastic
simulations of the metapopulation model (symbols) for $N=250$ (panel \textsf{\textbf{b}}) and $N=100$ (panels \textsf{\textbf{c}} and \textsf{\textbf{d}}) patches.
The dashed lines correspond to Levins' dynamics (\ref{eq:discr_lev}), discussed in the following section.

\section*{Levins' equation}\label{sec:le}
Points on the brink of extinction in Fig.~\ref{fig:2}{\panel a} (solid line) correspond to parameter values 
where the metapopulation dynamics \eqref{eq:detdyn1} suffers a qualitative change as mentioned in the preceding section. 
Slightly above this line, the metapopulation dynamics simplifies considerably, it becomes essentially one-dimensional \citep{GuH83,Dyk94,Eriksson2011}.
In this section we describe the form of this one-dimensional equation, show that it corresponds to a time-discrete version of Levins' equation
\eqref{eq:levins}, and determine the parameters $c$ and $e$ in this equation in terms of local
demographic parameters.

Our results are valid for metapopulations close to (and above) the critical line in Fig.~\ref{fig:2}{\panel a}.
We use the parameter
\begin{equation}\label{eq:delta}
        \delta = \frac{m - m_{\rm c}}{m_{\rm c}}\,,
\end{equation}
to characterise how close the population is to the brink of extinction ($\delta=0$).
For small values of $\delta$ the metapopulation dynamics \eqref{eq:detdyn1} 
exhibits a so-called \lq slow mode\rq: a linear combination of the $f_j$ that changes slowly in time.
The remaining \lq fast variables\rq{} (different linear combinations of the $f_j$) 
rapidly relax to steady states that depend on the instantaneous value of
the slow mode. We denote the slow mode by $Q_1$, it is given by a linear combination of the $f_j$:
\begin{equation}
Q_1(t) = \sum_j L_{1j} f_j(t)\,.
\end{equation}
\begin{figure}[tp]
   \centering
 \includegraphics[width=0.3\textwidth ,height=0.3\textwidth ]{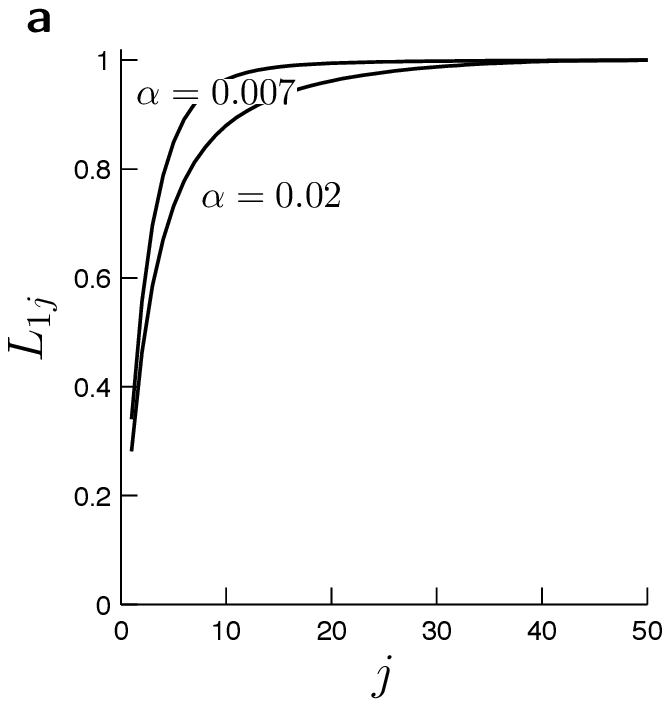}\qquad
 \includegraphics[width=0.3\textwidth ,height=0.3\textwidth ]{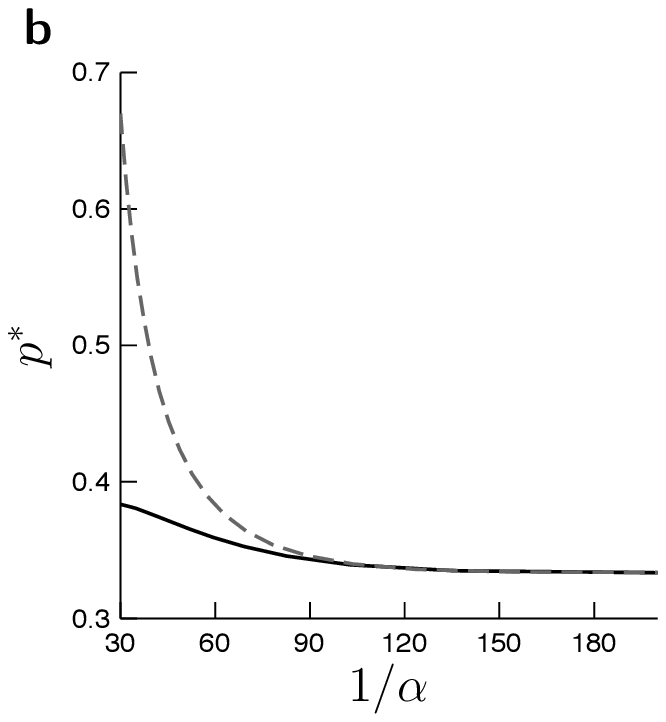}\qquad
   \includegraphics[width=0.3\textwidth,height=0.3\textwidth]{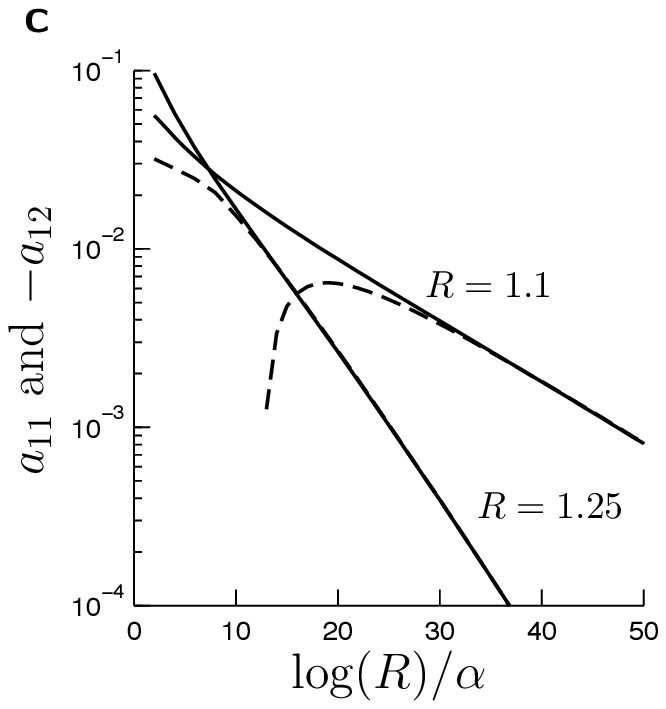}
   \caption{\label{fig:3}
(\textsf{\textbf{a}}) Shows the coefficients $L_{1j}$ as a function of $j$. The solid lines correspond to $R=1.25$ and the
values of $\alpha$ indicated in the figure (corresponding the parameters in Fig.~\ref{fig:2}).  (\textsf{\textbf{b}}) 
Shows the fraction $p^* = \sum_jf_j^*$ of occupied patches at the steady state as a function of $\alpha^{-1}$ for fixed $R=1.25$ and $\delta = 0.5$. Solid line corresponds to the deterministic dynamics, Eq. \eqref{eq:detdyn1}. 
The dashed line corresponds to Levins dynamics Eq.~\eqref{eq:discr_lev}, 
where we have interpreted $Q_1$ to be equal to the fraction of occupied patches. This is accurate only for small values of $\alpha$.
(\textsf{\textbf{c}}) 
Shows the coefficients $a_{11}$ and $-a_{12}$, Eqs.~(\ref{S-eq:c11_appB},\ref{S-eq:c11_appB2}) as a function of $\alpha^{-1}$ for the two values of $R$ indicated in the figure. Solid lines correspond to $a_{11}$, and dashed lines to $a_{12}$.  For convenience the $x$-axis is rescaled by the factor $\log(R)$.}
\end{figure}

In appendix~C we show how to compute the coefficients $L_{1j}$. The details do not matter, but it
is important to note the $L_{1j}\rightarrow 1$ in the limit of large values of $R$ and small values of $\alpha$. 
In other words, when the local dynamics is much faster than the dispersal dynamics, the slow mode corresponds to the fraction of occupied patches 
\begin{equation}
Q_1(t) \approx \sum_{j=1}^{\infty}f_j(t)\,.
\end{equation}
See Fig.~\ref{fig:3}(\textsf{\textbf{b}}). 
In this limit we find the following equation for the change of $Q_1$ from one
generation to the next:
\begin{equation}\label{eq:discr_lev}
	\Delta Q_{1}=  cQ_{1}\big(1-Q_{1}\big) - e Q_{1} \, ,
\end{equation}
This is a discrete version of Levins' equation (\ref{eq:levins}). 
The colonisation and extinction rates $c$ and $e$ in Eq.~(\ref{eq:discr_lev}) 
are derived in appendix~C. They depend on $R$ and $\alpha$ as well as on $\delta$:
\begin{equation}\label{eq:levrates}
	c = -a_{12}(1+\delta) \, \quad \text{and} \quad e = c - a_{11}\delta\, .
\end{equation}
The coefficients $a_{11}$ and $a_{12}$ depend on the values of $R$ and $\alpha$.
This dependence is illustrated in Fig.~\ref{fig:3}\panel{c}}. Eq.~(\ref{eq:levrates})
show how the rates $c$ and $e$ in Levins' equation (\ref{eq:discr_lev}) depend
upon the life history of the local populations (determined by $r$ and $K$ in the case
considered here, and upon the migration rate).

In panels {\panel b} to {\panel d} of Fig.~\ref{fig:2} we
compare Levins' dynamics (described by  Eqs.~(\ref{eq:discr_lev}) and (\ref{eq:levrates})), to results of direct simulations of the metapopulation model.
As pointed out above, and as discussed in appendix~C, we derived these equations assuming that the metapopulation
is on the brink of extinction (solid line in panel {\panel a}), and that the local dynamics is fast compared to migration. 
Both assumptions are satisfied in panel {\panel d}, and we observe excellent agreement between Eqs.~(\ref{eq:discr_lev},\ref{eq:levrates})
and the results of direct stochastic simulations, except for a rapid initial transient.
In panel {\panel b}, the agreement is not as good. This is due to the fact that the parameter
$\alpha$ is too large, implying that the local and global time scales are not as clearly separated
as in panel {\panel d}. Panel {\panel c} corresponds to parameter values much further
from the critical line. In this case, Eqs.~(\ref{eq:discr_lev},\ref{eq:levrates}) fail to 
describe the metapopulation dynamics.
\begin{figure}[t]
   \includegraphics[width=0.4\textwidth,height=0.4\textwidth]{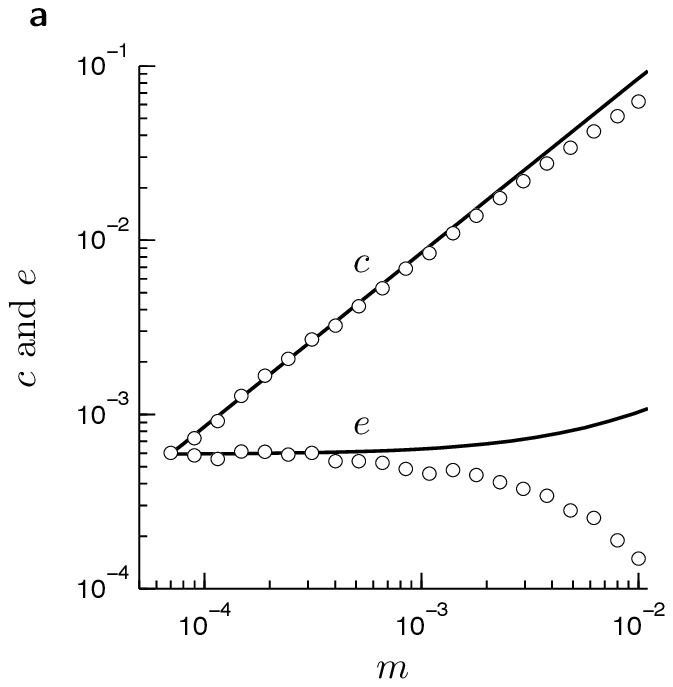}\qquad
   \includegraphics[width=0.4\textwidth,height=0.4\textwidth]{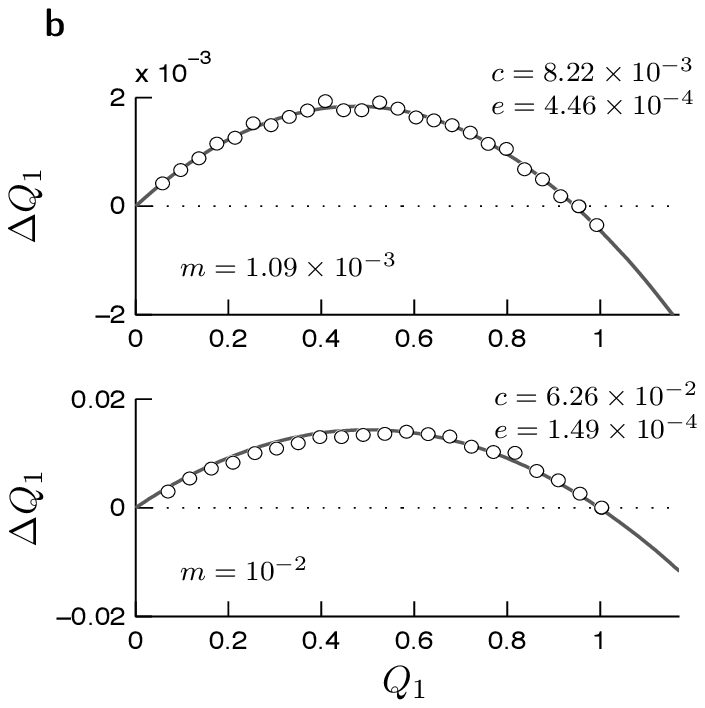}
   \caption{\label{fig:4} (\textsf{\textbf{a}})
Colonisation and extinction rates  $c$ and $e$ in Levins' equation (\ref{eq:discr_lev}). The solid lines correspond to the theoretical values, Eq.~\eqref{eq:levrates}, symbols ($\bigcirc$) to direct stochastic simulations for $N=250$, averaged over 100 stochastic realisations. The procedure to calculate $e$ and $c$ from the simulations is described in section \ref{sec:le}. (\textsf{\textbf{b}}) $\Delta Q_1$ as a function of $Q_1$ for the two values of the emigration rate $m$ indicated in each panel. Symbols ($\bigcirc$) correspond to direct stochastic simulations for $N=250$ averaged over 100 stochastic realisations. Solid lines correspond to least-squares fits of the simulation results to
Eq.~\eqref{eq:discr_lev}. Each panel shows the resulting values of $e$ and $c$. Parameters for both panels are $R=1.25$ and $\alpha=0.008$.  }
\end{figure}

The dependence of the colonisation and extinction rates on the emigration rate $m$ is shown in Fig.~\ref{fig:4}{\panel a}.
Solid lines correspond to the analytical result (\ref{eq:levrates}) and (\ref{S-eq:c11_appB},\ref{S-eq:c11_appB2}) in appendix~C.  
The results of direct stochastic simulations (symbols) were obtained as follows.
First, for each set of parameters we determined the dynamics of $\ve f(t)$ by direct stochastic simulations.
Second, assuming a clear time-scale separation between the local patch dynamics and migration, the resulting $\ve f(t)$ was projected according to $Q_{1}(t) \approx \sum_i f_{i}(t)$. 
Third, the resulting values of $Q_{1}(t)$ were averaged over  100 independent simulations. 
Fourth, from this average, we computed $\Delta Q_1$ as a function of $Q_1$, 
and found the coefficients of Eq.~\eqref{eq:discr_lev} by a least-squares polynomial fit 
to the parabola on the right-hand side of Eq.~\eqref{eq:discr_lev}.
Fits are shown in panel {\panel b}. 

As the migration rate approaches its critical value, Fig.~\ref{fig:4}{\panel a} shows that the colonisation rate approaches the same value as the extinction rate in agreement with Eq.~(\ref{eq:levrates}) for $\delta \to 0$.
For larger values of $m$, the extinction rate is approximately independent of $m$.
This is a consequence of the fact that $a_{11} \approx -a_{12}$ in the limit
of time-scale separation (see Fig.~\ref{fig:3}\textsf{\textbf{c}}). The colonisation rate is seen to scale as $\sim 1+\delta$,
as predicted by Eq.~(\ref{eq:levrates}).
As the emigration rate increases further, we observe deviations
between the theoretical prediction Eq.~(\ref{eq:levrates}) and the numerical results.
This is expected: the approximations leading to Eqs.~(\ref{eq:discr_lev},\ref{eq:levrates})
are valid on the brink of extinction and fail further away from the critical migration line.
 
\section*{Discussion} 
\label{sec:conc}
In this paper we have analysed the dynamics of a stochastic metapopulation model 
based on coupled Ricker maps, and have shown how Levins' dynamics emerges 
on the brink of extinction. Our derivation of a discrete version of Levins' model is valid in the 
limit of many patches, and when the local dynamics is faster than the global migration dynamics. Levins' model 
emerges on the brink of extinction because the deterministic dynamics (valid in the limit of many patches)  exhibits a unique slow mode. 
Our finding that Levins' equation is applicable at the brink of extinction, 
based on an analytical study of a stochastic metapopulation model, 
is in line with results obtained by other authors who have attempted 
to bridge the gap between the results of direct numerical simulations of 
individual-based stochastic metapopulation models and Levins' model. Keeling \citet{Keeling2002} 
did so by fitting results from numerical simulations to the rates $c$ and $e$ 
in Levins' model. Fronhofer {\em et al.} \citet{Fronhofer2012}, also through numerical simulations, 
have explored under which circumstances individual-based models qualitatively behave as metapopulations. 
In this paper, we provide a convenient analytical framework that allows us to define formally under which circumstances 
Levins' model provides an adequate approximation of the full metapopulation dynamics. 

Another important outcome of our analytical treatment of stochastic metapopulations 
is our derivation of Levins' key parameters $c$ and $e$ in terms of life history parameters 
from the Ricker dynamics that describes the local populations, see Eq.~(\ref{eq:levrates}) in the main text and Eqs.~(\ref{S-eq:c11_appB}), (\ref{S-eq:c11_appB2}) in appendix~C.
A numerical implementation of our method for computing $c$ and $e$ is available in the supplementary 
material.

While Levins' model has been used extensively by ecologists to explore the behaviour of metapopulations, 
it is very difficult to fit the model to ecological data. 
Ideally, one would need long time series tracking individual populations and recording extinction and 
colonisation. However, such datasets are rare, and unlikely to be available for species of management 
interest because of their risk of global extinction. Some of those rare examples are presented in Refs. \citep{Hanski1995,Hanski1994}, where the authors have additionally fitted the ecological data in question to a metapopulation model in order to use it predictively with some success. 
In this paper, we have provided an alternative solution to this problem, as 
demographic parameters are often easily available from short term studies of individual populations, especially for larger mammals in which population turnover 
is likely to be low \citep{Elmhagen2001}. 
In addition, since the parameters in Levins' equation depend upon the local population dynamics, as well as on the dispersal between patches, 
parameters obtained by fitting to a given data set do not necessarily transfer to other contexts.
In contrast,  our mechanistic connection between local population dynamics and the coefficients of Levins' equation allows using a fitted model to investigate which aspects of the local population dynamics has the strongest effect on extinction and colonisation rates. For example, is the population most sensitive to adult or juvenile mortality, or is it mainly affected by fecundity?

We have focused on the Ricker map, which was originally devised for fish \citep{Ricker1954},
but is also applicable to a variety of taxa, such as mammals \citep{Sherratt2000}, birds \citep{Holyoak1996,Wiklund2001},
and insects \citep{Barlow2002,Melbourne2008}. Furthermore, the methods used in this paper are not restricted to the Ricker map, 
but can be used to obtain estimates of $c$ and $e$ based on any other type of local stochastic population dynamics, as long as dispersal is still random and uniform (but the number of dispersing individuals can have any dependence on the local population size). The ability to fit Levins' model
to metapopulations at the brink of extinction should provide a potentially powerful tool for applied ecologists,
who can now, with minimal information about the species of interest, leverage several decades of
detailed analysis of the dynamics of Levins' model.

\section*{Acknowledgments}
Financial support by Vetenskapsr\aa{}det, by the G\"oran Gustafsson Foundation
for Research in Natural Sciences and Medicine, and by CeMEB  are gratefully acknowledged.
FEW wishes to thank the Centro Internacional de Ciencias in Cuernavaca for its hospitality during an extended stay.

\appendix
\section*{Appendix A}
\label{app:newA}

\renewcommand{\theequation}{A.\arabic{equation}}
\setcounter{equation}{0}

In this appendix we derive an expression for $P_{j \to i}(t)$ in Eq. \eqref{eq:detdyn1}. We calculate $P_{j \to i}(t)$ as follows. 
First consider reproduction and survival to adulthood. In our model the number of individuals after reproduction is Poisson distributed with mean $R \eta_t$.
Survival is a binomial process with survival probability $\exp(-\alpha \eta_t)$. Therefore the number of survivors 
is Poisson distributed with expected value
$R \eta_t \ee^{-\alpha \eta_t}$. After dispersing individuals have left the patch (but before any dispersers arrive), 
the number of individuals is Poisson-distributed with expected value  $(1-m)R \eta_t \ee^{-\alpha \eta_t}$.
Lastly, since dispersing individuals choose their destination independently,
the number of individuals dispersing into a given patch is Poisson-distributed with expected value
\begin{equation}
\label{eq:It}
D(t) = m \sum_{j=0}^{\infty} R j \exp(-\alpha j) f_{j}(t)\,.
\end{equation}
Bringing it all together, the probability $P_{j \to i}(t)$ 
in Eq.~\eqref{eq:detdyn1} of finding $i$ individuals in the patch in the next generation starting from $j$ individuals in the 
current generation is
\begin{equation}\label{eq:detdynbin}
P_{j \to i}(t) = \frac{\left[(1-m) j \ee^{-\alpha j} R + D(t)\right]^i}{i!}\ee^{-(1-m) j \ee^{-\alpha j} R - D(t)}\,.
\end{equation} 
We took advantage of the fact that the number of resident individuals and the number of dispersing individuals arriving at the patch are two independent
Poisson-distributed numbers (the sum of which is also Poisson distributed).
Note that $P_{j \to i}(t)$ depends upon $f_{i}(t)$ through the number of immigrants, $D(t)$. 
This non-linearity is necessary for a quasi-stable steady state to exist.
Note also that the components $f_{i}(t)$ are not independent, because the number of patches is equal to $N$.
We incorporate this constraint explicitly by substituting $1 - \sum_{i=1}^{\infty}f_{i}(t)$ for $f_{0}(t)$.
This results in the following dynamics for $\ve f(t) \equiv (f_{1}(t), f_{2}(t), \dotsc)$
\begin{equation}\label{eq:discr_constr} 
f_{i}(t+1) = \sum_{j=1}^\infty \frac{\big[(1-m) j \ee^{-\alpha j} R + D(t)\big]^i}{i!}\ee^{-(1-m) j \ee^{-\alpha j} R - D(t)} f_{j}(t) 
+ \frac{D(t)^i}{i!}\ee^{- D(t)} \Big( 1 - \sum_{j=1}^\infty f_{j}(t) \Big)\,.
\end{equation}

\section*{Appendix B}
\label{app:B}

\renewcommand{\theequation}{B.\arabic{equation}}
\setcounter{equation}{0}

In order to determine how $m_{\rm c}$ depends upon the life history parameters
$R$ and $\alpha$, we investigate the linear stability of the steady state $\ve f^\ast$ by determining the eigenvalues
of the stability matrix 
\begin{equation}
{\bf A} = \left. \frac{\partial \ve{f}(t+1)}{\partial \ve{f}({t})} \right|_{\ve f = \ve f^\ast}\, .
\end{equation}
The state $\ve f^\ast$ is linearly stable provided all eigenvalues $\lambda_j$
are less than unity in modulus.  The bifurcation in the dynamics, that is, the point where the extinction steady state $\ve f = \ven 0$ changes from being unstable to being stable, occurs
when the leading eigenvalue (that is, the largest in absolute value) approaches
unity: $\lambda_1(m) \rightarrow 1$ as $m\rightarrow m_{\rm c}$.
This criterion allows us to compute the dependence of $m_{\rm c}$ upon $\alpha$. We do this numerically in the following manner. At the bifurcation we have $\ve f^* = \ven 0$ and the matrix $\ma A = \ma A^{(0)}$, the elements of which is given in appendix D. We then numerically compute the eigenvalues of $\ma A^{(0)}$ for a suitable range of $m$, thus constructing a function $\lambda_1(m)$. Finally, from this function we interpolate the value of $m=m_{\rm c}$ such that $\lambda_1(m_{\rm c}) = 1$.

\section*{Appendix C}
\label{app:C}

\renewcommand{\theequation}{C.\arabic{equation}}
\setcounter{equation}{0}

In this section we show how Levins' dynamics  Eqs.~(\ref{eq:discr_lev},\ref{eq:levrates}) emerges
from the deterministic dynamics  (\ref{eq:discr_constr}).
In order to show how the dynamics in Eq.~\eqref{eq:discr_constr} simplifies for small positive values of $\delta$, 
we follow Refs.~\citet{GuH83,Dyk94,Eriksson2011} and expand the right-hand side of Eq.~\eqref{eq:discr_constr} 
in powers of $\delta$, noting that the components of $\ve f(t)$ are of order $\delta$.  
To simplify the notation we define $\Delta f_{i}(t+1) = f_{i}(t+1)-f_{i}(t)$ and drop the generation dependence $t$:
\begin{align}
\label{eq:expand3}
\Delta f_i  \approx {} & -f_i + \sum_j A_{ij}^{(0)} f_j + \delta \sum_j A_{ij}^{(1)} f_j  + \frac{1}{2}(1+\delta) \sum_{jk} A_{ijk}^{(2)}  f_j f_k  \nn \\
& + \frac{1}{2} \delta \sum_{jk} A_{ijk}^{(3)}  f_j f_k + \frac{1}{3!}\sum_{jkl} A_{ijkl}^{(4)}  f_j f_k f_l +\cdots\,.
\end{align}
Here $A_{ij}^{(0)}$ are the elements of the stability matrix ${\bf A}$ evaluated at $\delta=0$.
The coefficients $A_{ij}^{(0)}$, $A_{ij}^{(1)}$, $A_{ijk}^{(2)}$, $A^{(3)}_{ijk}$ and $A^{(4)}_{ijkl}$ are given in appendix D.

In order to identify the slow mode, we diagonalise 
the linearisation ${\bf A}^{(0)}$ of the deterministic dynamics (\ref{eq:discr_constr}) at $\delta=0$:
\begin{equation}
\label{eq:lambda}
{\bf A}^{(0)} \ve R_\beta = \lambda_\beta \ve R_\beta\,,\quad \ve L_\beta \tr {\bf A}^{(0)} = \ve L_\beta \tr \lambda_\beta\,,\quad
\mbox{for $\beta = 1,2,\ldots$}.
\end{equation}
We take the left and right eigenvectors to be bi-orthogonal and ordered as 
\begin{equation}
\label{eq:LR}
	\ve L_\beta  \tr  \ma{A}^{(0)} \ve R_\gamma= \lambda_\beta \delta_{\beta\gamma}\, , \quad |\lambda_1| > |\lambda_2| > |\lambda_3| > \dotsb \, . 
\end{equation}
The condition (\ref{eq:LR}) does not determine the normalisation of the eigenvectors. 
Fig.~\ref{fig:3}(\textsf{\textbf{a}}) shows that the components $L_{1j}$ approach a $j$-independent constant.
We take this constant to be unity, thus fixing the normalisation of the leading left
eigenvector, and thus by orthonormality fixing also the normalisation of $\ve R_1$.
The limiting value is approached more rapidly for larger values of $R$ and smaller values of $\alpha$. 

At $\delta=0$ we have that $\lambda_1=1$. For small values of $\delta$ we find  that $\lambda_1$ is real and slightly larger than unity,
$\lambda_1 = 1 + O(\delta)$, and  $|\lambda_\beta| < 1$ for $\beta> 1$. Thus, for small values of $\delta$,
 perturbations away from $\ve f = \ven 0$ 
grow slowly in the direction of $\ve L_1$, and decay rapidly in the directions $\ve L_{\beta}$, $\beta > 1$. 
Thus, the slow mode is given by
\begin{align}\label{eq:slowmode}
	Q_1 = {}& \ve L_1 \tr  \ve f\, .
\end{align}
This equation implies that the slow mode corresponds to the fraction of occupied patches if
$\ve L_1\tr = (1,1,1,\dotsc)$. Figure~\ref{fig:3}\panel{a} shows the components of $\ve L_1$ for different
values of $\alpha$. We adopted a normalisation convention leading to $\ve L_1\tr \rightarrow (1,1,1,\dotsc)$ as $R$ becomes
large and $\alpha$ tends to zero. This ensures that $Q_1$ can be interpreted as the fraction of occupied patches in this limit 
as shown in Fig.~\ref{fig:3}\panel{c}.
The fast variables are given by
\begin{align}
\label{eq:fast}
	Q_{\beta} = {}& \ve L_{\beta} \tr \ve f \quad \text{for $\beta > 1$.}
\end{align}
The fast variables quickly approach local equilibria that depend on the instantaneous value of the slow variable, $Q_1$. 
In order to derive an equation for the dynamics of $Q_1$ we start from (\ref{eq:expand3}). 
Inserting (\ref{eq:slowmode}) and (\ref{eq:fast}) into (\ref{eq:expand3}) we find an expansion
of $\Delta Q_\beta$ in powers of $\delta$.  We expect that $Q_1$ is of order $\delta$.
Let us first consider this expansion to second order in $\delta$:
\begin{align}
\label{eq:expand2}
\Delta Q_\beta \approx{}& -Q_{\beta} + \ve L_\beta\tr {\bf A}^{(0)} \sum_\gamma \ve R_\gamma Q_\gamma
        + \delta  \sum_{ij\gamma} L_{\beta i} A_{ij}^{(1)} R_{\gamma j} Q_\gamma \nn \\
        &{}+\frac{1}{2} \sum_{ijk\mu\nu} L_{\beta i} A_{ijk}^{(2)}  R_{\mu j} R_{\nu k} Q_\mu Q_\nu+O(\delta^3)\,.
\end{align}
The fast variables quickly approach local equilibria  given by $\Delta Q_\beta = 0$. This yields
\begin{equation}
\label{eq:Qa}
 Q_\beta \approx \frac{a_{\beta 1}\delta Q_1 + a_{\beta 2}  Q_1^2}{1 - \lambda_\beta}\quad\mbox{for $\beta > 1$}
\end{equation}
On the right-hand side of Eq.~(\ref{eq:expand2}), terms involving the fast variables are neglected. 
This is consistent since {Eq.~\eqref{eq:Qa}} implies that $Q_\beta$ are of order $\delta^2$ close to the critical line.
The coefficients $a_{\beta 1}$ and $a_{\beta 2}$ are given by
\begin{align}
\label{S-eq:c11_appB}
a_{\beta 1}={}&  \sum_{ij} L_{\beta i} A_{ij}^{(1)} R_{1j} \,,\\
a_{\beta 2}={}&\frac{1}{2} \sum_{ijk} L_{\beta i} A_{ijk}^{(2)} R_{1j} R_{1k} \,.
\label{S-eq:c11_appB2}
\end{align}
Figure \ref{fig:3}\panel{c} shows how $a_{11}$ and $a_{12}$ depend on $R$ and $\alpha$.

We now make a further simplification, assuming time-scale separation between the fast local processes and slow emigration.
In keeping with this approximation we neglect terms of order $O(m_{\rm c}^2)$ and higher. 
To orders $\delta^3$ and $m_{\rm c}$ we find from Eqs.~(\ref{eq:expand3}), (\ref{eq:slowmode}), and the
results of appendix D that
\begin{equation}\label{eq:discr_3rd}
	\Delta Q_1 =  a_{11}\delta \,Q_1 + a_{12}(1 + \delta)Q_{1}^2 + O(\delta^4) + O(m_{\rm c}^2)\,.
\end{equation}
Eq.~\eqref{eq:discr_3rd} has the form of Levins' equation (for discrete time evolution):
\begin{equation}
\label{eq:levinsB}
	\Delta Q_{1}=  cQ_{1}\big(1-Q_{1}\big) - e Q_{1} \, .
\end{equation}
The colonisation and extinction rates are given by:
\begin{equation}\label{eq:levrates_appB}
	c = -a_{12}(1+\delta) \, \quad \text{and} \quad e = c - a_{11}\delta\, .
\end{equation}
We note that $a_{11}$ is positive while $a_{12}$ is negative. Moreover, as $R$ becomes large and $\alpha \to 0$, 
the coefficient $a_{11}$ is approximately equal to $-a_{12}$. This implies that the coefficient $e$ is approximately independent of $\delta$, 
while $c$ scales as $1+\delta$, see Fig.~\ref{fig:4}{\panel a} in the main text. 
A numerical implementation of our method for computing $c$ and $e$ is available as a MATLAB script in the supplementary material.
The fixed points of Eq.~\eqref{eq:levinsB} are the extinction point $0$ and the steady-state $Q_1^{*} = 1- e/c$.
For $|c-e| < 2$ and $c<e$, the extinction point $0$ is stable and the steady state $Q_1^{*}$ is unstable. At $c=e$, both fixed points 
meet and exchange their stability for $c>e$. 

At $|c-e| = 2$ a period-doubling bifurcation occurs \citep{Ott}, and as $|c-e|$ increases still further, chaos ensues. 
See \citep{Hastings1994} for the study of chaotic transients.
However, this behaviour is outside the range of applicability of Eq.~\eqref{eq:levinsB}, that is, too far from the critical line in Fig.~\ref{fig:2}{\panel a}.

We conclude by summarising the three assumptions that were necessary to obtain Eqs.~(\ref{eq:levinsB}) and (\ref{eq:levrates_appB}).
First, it was assumed that the number $N$ of patches is large. Second, it was assumed that the metapopulation
is on the brink of extinction (allowing us to assume that the parameter $\delta$ is small). Third, it was assumed
that the local dynamics is much faster than the global dynamics (allowing us to assume that $m_{\rm c}$
is small).

\newpage
\section*{Appendix D}
\label{app:A}

\renewcommand{\theequation}{D.\arabic{equation}}
\setcounter{equation}{0}

In this section we give the coefficients occurring in the expansion (\ref{eq:expand3}): the matrix elements of  ${\ma A}^{(0)}$ and ${\ma A}^{(1)}$,
as well as the coefficients $A_{ijk}^{(2)}$, ${A}_{ijk}^{(3)}$ and ${A}_{ijkl}^{(4)}$.
All coefficients are evaluated at $\delta=0$, where the steady state is $\ve f = \ven 0$.
The stability matrix ${\ma A}^{(0)}$ has elements
\begin{alignat}{2}
	A_{1j}^{(0)} =	{} & { \rho_j \ee^{-\rho_j}  + m_{\rm c} j \ee^{-\alpha j} R \,,} \nn \\ 
	A_{ij}^{(0)} = {} &  {\frac{\rho_j^i}{i!} \ee^{-\rho_j}\,,}  \qquad && \text{for $i > 1$\,,} 
\end{alignat}
with $\rho_j = (1-m_{\rm c}) j \ee^{-\alpha j} R$. The elements of the matrix ${\ma A}^{(1)}$ are given by 
\begin{alignat}{2}
	A_{1j}^{(1)} =	{} & {m_{\rm c} j R \exp[{-\alpha j} {-\rho_j }] 
	\big( \rho_j  - 1  \big)  +  m_{\rm c}j \ee^{-\alpha j} R \,,} \nn \\ 
	A_{ij}^{(1)} = {} &  { m_{\rm c} j R \exp[{-\alpha j} {-\rho_j }]}  
	\sum_{k=i-1}^i \frac{ {\rho_j}^k}{k!}(-1)^{i-k} 
	  \,,  \qquad &&  \text{for $i > 1$.} 
\end{alignat}
The coefficients ${A}_{ijk}^{(2)}$ are given by
{\allowdisplaybreaks
\begin{alignat}{2}
	A_{1jk}^{(2)} = {} & { m_{\rm c} k R \exp[{-\alpha k - \rho_j } ]
	\left( 1	- \rho_j  \right)} 
	 + m_{\rm c} j R \exp[-\alpha j - \rho_k ]
	\left( 1 - \rho_k  \right) \nn \\
	& {-  m_{\rm c} R \big( j \exp({-\alpha j}) + k \exp({-\alpha k}) + 2  m_{\rm c} R j k \exp(-\alpha(j+k))\big)\,, }\nn \\
	A_{2jk}^{(2)} = {} & { m_{\rm c} k R \exp[{-\alpha k - \rho_j } ]
	\left( \rho_j  
	- \frac{{\rho_j}^2}{2}  \right) }
	 {+  m_{\rm c} j R \exp[{-\alpha j - \rho_k } ]
	\left( \rho_k 
	- \frac{{\rho_k}^2}{2}  \right) }\nn \\
	& {+ ( m_{\rm c} R)^2 j k \exp(-\alpha(j+k))\,, }\nn \\
	A_{ijk}^{(2)} = {} & { m_{\rm c} k R \exp[{-\alpha k - \rho_j } ]
	\sum_{\nu=i-1}^i \frac{ {\rho_j}^\nu}{\nu!}(-1)^{\nu-i+1} }\nn \\
	& {+  m_{\rm c} j R \exp[{-\alpha j - \rho_k } ]
	\sum_{\nu=i-1}^i \frac{ {\rho_k}^\nu}{\nu!}(-1)^{\nu-i+1}\,, } && \hspace{-42pt} \text{for $i > 2$.} 
\end{alignat}  
}
The coefficients ${A}_{ijk}^{(3)}$
are given by
\begin{alignat}{2}
	A_{1jk}^{(3)} = {} &  m_{\rm c}^2 k j R^2 \exp[{-\alpha (k+j) - \rho_j } ] 
	\left( 2- \rho_j \right) 
	 + m_{\rm c}^2 k j R^2 \exp[{-\alpha (k+j) - \rho_k } ] 
	\left( 2- \rho_k \right) \nn \\
	&-2 m_{\rm c}^2 R^2 j k \exp(-\alpha(j+k))\, , \nn \\
	A_{2jk}^{(3)} = {} &  m_{\rm c}^2 k j R^2 \exp[{-\alpha (k+j) - \rho_j } ] 
	\left( -1 + 2 \rho_j  
	- \frac{{\rho_j}^2}{2}  \right) \nn \\
	& + m_{\rm c}^2 k j R^2 \exp[{-\alpha (k+j) - \rho_k } ]
	\left( -1 +2 \rho_k
	- \frac{{\rho_k}^2}{2}  \right) \nn \\
	&+2 m_{\rm c}^2 R^2 j k \exp(-\alpha(j+k))\, , \nn \\
	A_{ijk}^{(3)} = {} &  m_{\rm c}^2 k j R^2 \exp[{-\alpha (k+j) - \rho_j } ]
	\left( -\frac{{\rho_j}^{i-2}}{(i-2)!}
	+2\frac{{\rho_j}^{i-1}}{(i-1)!}
	- \frac{{\rho_j}^i}{i!}  \right) \nn \\
	& + m_{\rm c}^2 k j R^2 \exp[{-\alpha (k+j) - \rho_k } ]
	\left( -\frac{{\rho_k}^{i-2}}{(i-2)!}
	+2\frac{{\rho_k}^{i-1}}{(i-1)!}
	- \frac{{\rho_k}^i}{i!}  \right)\, , \nn \\
	& {}&& \hspace{-42pt} \text{for $i > 2$.} 
\end{alignat}
Finally, the coefficients ${A}_{ijkl}^{(4)}$ are given by
{\allowdisplaybreaks
\begin{alignat}{2}
	A_{1jkl}^{(4)} = {} &  m_{\rm c}^2 k l R^2 \exp[{-\alpha (k+l) - \rho_j } ] 
	\left( -2 + \rho_j  \right) 
	 +  m_{\rm c}^2 j k R^2 \exp[{-\alpha (j+k) - \rho_l } ] 
	\left( -2 + \rho_l   \right) \nn \\
	& +  m_{\rm c}^2 j l R^2 \exp[{-\alpha (j+l) - \rho_k } ] 
	\left( -2 + \rho_k   \right)  \nn \\
	& - 2 m_{\rm c}^2 R^2 \Big( j k \exp[-\alpha (j+k)] + jl \exp[-\alpha (j+l)] + kl \exp[-\alpha (k+l)]\Big) \nn \\
	& + 3 m_{\rm c}^3 R^3 jkl \exp[-\alpha (j+k+l)]\, , \nn \\  
	A_{2jkl}^{(4)} = {} &  m_{\rm c}^2 k l R^2 \exp[{-\alpha (k+l) - \rho_j } ]
	 \left( 1 -2 \rho_j  
	+ \frac{{\rho_j}^2}{2}  \right) \nn \\
	&+  m_{\rm c}^2 j k R^2 \exp[{-\alpha (j+k) - \rho_l } ]
	 \left( 1 -2 \rho_l  
	+ \frac{{\rho_l}^2}{2}  \right) \nn \\
	&+  m_{\rm c}^2 j l R^2 \exp[{-\alpha (j+l) - \rho_k } ]
	 \left( 1 - 2 \rho_k  
	+ \frac{{\rho_k}^2}{2}  \right) \nn \\
	& + m_{\rm c}^2 R^2 \Big( j k \exp[-\alpha (j+k)] + jl \exp[-\alpha (j+l)] + kl \exp[-\alpha (k+l)]\Big) \nn \\
	& - 3 m_{\rm c}^3 R^3 jkl \exp[-\alpha (j+k+l)]\, , \nn \\  
	A_{3jkl}^{(4)} = {} &  m_{\rm c}^2 k l R^2 \exp[{-\alpha (k+l) - (1-m_{\rm c}) j \ee^{-\alpha j} R } ]
	 \left( \rho_j  
	-2\frac{{\rho_j}^{2}}{2}
	+ \frac{{\rho_j}^3}{3!}  \right) \nn \\
	&+  m_{\rm c}^2 j k R^2 \exp[{-\alpha (j+k) - \rho_l } ] 
	 \left( \rho_l  
	-2\frac{{\rho_l}^{2}}{2}
	+ \frac{{\rho_l}^3}{3!}  \right) \nn \\
	&+  m_{\rm c}^2 j l R^2 \exp[{-\alpha (j+l) - \rho_k } ] 
	 \left( \rho_k 
	-2\frac{{\rho_k}^{2}}{2}
	+ \frac{{\rho_k}^3}{3!}  \right) \nn \\
	& + m_{\rm c}^3 R^3 jkl \exp[-\alpha (j+k+l)]\, , \nn \\  
	A_{ijkl}^{(4)} = {} &  m_{\rm c}^2 k l R^2 \exp[{-\alpha (k+l) - \rho_j } ]
	 \left( \frac{{\rho_j}^{i-2}}{(i-2)!}
	-2\frac{{\rho_j}^{i-1}}{(i-1)!}
	+ \frac{{\rho_j}^i}{i!}  \right) \nn \\
	&+  m_{\rm c}^2 j k R^2 \exp[{-\alpha (j+k) - \rho_l } ]
	 \left( \frac{{\rho_l}^{i-2}}{(i-2)!}
	-2\frac{{\rho_l}^{i-1}}{(i-1)!}
	+ \frac{{\rho_l}^i}{i!}  \right) \nn \\
	&+  m_{\rm c}^2 j l R^2 \exp[{-\alpha (j+l) - \rho_k } ] 
	 \left( \frac{{\rho_k}^{i-2}}{(i-2)!}
	-2\frac{{\rho_k}^{i-1}}{(i-1)!}
	+ \frac{{\rho_k}^i}{i!}  \right) \, , 
	 && \hspace{-40pt} \text{for $i > 3$.} 
\end{alignat}  
}



\newpage 

\begin{figure}[htp]   
   \includegraphics[width=1.0\textwidth]{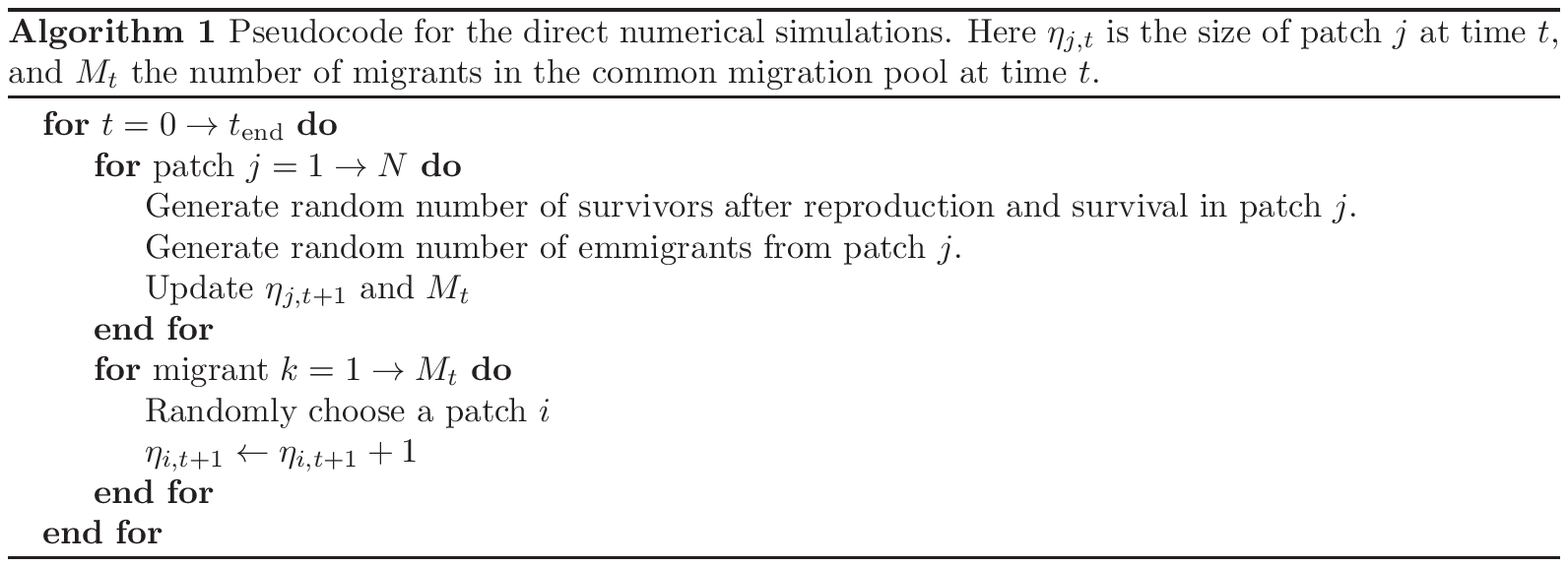}
\end{figure}


\end{document}